\documentclass{llncs}

\usepackage{graphicx}
\usepackage{tabularx}
\usepackage{hyperref}
\usepackage{graphicx}
\usepackage[british]{babel}
\usepackage{eurosym}
\usepackage{pbox}
\usepackage{url}
\usepackage{xspace}
\usepackage{lipsum}
\makeatletter
\g@addto@macro{\UrlBreaks}{\UrlOrds}
\usepackage{booktabs}
\usepackage[misc]{ifsym}
\usepackage[font=small,labelfont=bf,labelsep=space]{caption}
\newcommand\blfootnote[1]{%
	\begingroup
	\renewcommand\thefootnote{}\footnote{#1}%
	\addtocounter{footnote}{-1}%
	\endgroup
}

\usepackage{geometry} 

\begin{document}
\mainmatter

\title{The Group Element of Cybercrime: Types, Dynamics, and Criminal Operations}
\titlerunning{The Group Element of Cybercrime}

\author{Jason R. C. Nurse\inst{1} \and
	Maria Bada\inst{2}}
\authorrunning{Nurse \& Bada}
%
\institute{School of Computing, University of Kent, UK\\
	\email{j.r.c.nurse@kent.ac.uk} \and
	Global Cyber Security Capacity Centre, University of Oxford, UK\\
	\email{maria.bada@cs.ox.ac.uk}}

\maketitle

\begin{abstract}
While cybercrime can often be an individual activity pursued by lone hackers, it has increasingly grown into a group activity, with networks across the world. This chapter critically examines the group element of cybercrime from several perspectives. It identifies the platforms that online groups—cybercriminal and otherwise—use to interact, and considers groups as both perpetrators and victims of cybercrime. A key novelty is the discovery of new types of online groups whose collective actions border on criminality. The chapter also analyzes how online cybercrime groups form, organize, and operate. It explores issues such as trust, motives, and means, and draws on several poignant examples, from Anonymous to LulzSec, to illustrate the arguments.\blfootnote{This is an article pre-print of the chapter: ``The Group Element of Cybercrime: Types, Dynamics, and Criminal Operations'' by Jason R.C. Nurse and Maria Bada, due to appear in The Oxford Handbook of Cyberpsychology (2018/19), Edited by Alison Attrill-Smith, Chris Fullwood, Melanie Keep, and Daria J. Kuss. https://dx.doi.org/10.1093/oxfordhb/9780198812746.013.36.}

\end{abstract} 

\section{Introduction}
\label{sec:introduction}
There are various perspectives through which cybercrime and its association with online groups can be studied, e.g., the groups that are responsible for cyber-attacks and similar acts of online aggression or the groups of individuals that are targeted. Anonymous is one of the most well-known of the hacker groups and has been linked to numerous high-profile online attacks. These include cyber-attacks on the FBI, US Department of Justice, and US Copyright Office~\cite{peckham2012sites}, declarations of war on banks and stock exchange markets~\cite{schwartzmj2016bank}, and more recent calls to action against U.S. President Donald Trump \cite{griffin2017anon}. Other popular cybercriminal groups are Lizard Squad, a group that forced the Sony’s PlayStation Network offline and caused a flight disruption with a bomb scare \cite{zorabedian2014lizar}; and the hacker group, Lulzsec, which stole private data from 24.6 million customers via a hack on Sony’s PlayStation Network \cite{arthur2013lul}.

In addition to these hacker groups, traditional organized crime groups are quickly expanding their presence into cyberspace. This is undoubtedly linked to the low barriers of entry, opportunity to vastly expand operations, and the perceived anonymity that the Internet provides. There is also a range of ad hoc groups consisting of members of the public who form online in support of a cause, but whose actions may be regarded as potentially criminal, e.g., the recent call to protest against Trump’s January 2017 inauguration with a distributed denial-of-service attack (DDoS) \cite{metzger2017ddos}. This protest campaign was publicized online and requested that the public flood the WhiteHouse.gov website with requests to ``demonstrate the will of the American people.'' Though the protest was later canceled because of the potential legal ramifications, it demonstrates the power of group action online.

Research has studied cybercriminal groups to varying extents to better understand their motivations, how they form and organize, and their techniques of attack \cite{choo2008criminal,olson2013we,mcguire2012organised}. Traditional organized crime groups, for example, are often driven online by the ability to up-skill quickly (via purchasing cybercrime services and tools) and therefore, to launch high-tech crimes with limited understanding and expertise \cite{aikenmp2014}. Online hacker groups are particularly interesting because unlike traditional groups, they may typically have to self-organize, i.e., as there may not be an agreed leader of the group to direct and co-ordinate operations, the group itself is responsible for these activities. Moreover, their actions are often based on causes, some of which the public may consider to be noble, and thus socially acceptable—the launch of \#OpISIS, a cyber-attack campaign against the ISIS terrorist network after the 2015 Paris attacks is one example. In this case, Anonymous even posted a video declaring (cyber) war on the Islamic State group---targeting their websites and social media accounts---in response to attacks such as the Charlie Hebdo massacre in Paris in January 2015 that killed twelve people.

Another perspective in the study of the group element of cybercrime is a focus on groups as the target of crimes. Young Internet users, for instance, are often studied as they represent a particularly vulnerable group online \cite{kowalski2014bullying}. Religion, as one might imagine, is also a topic that has resulted in numerous crimes online, particularly harassment and hate speech. Race is another polarizing subject in online communities, with many online hate groups actively congregating to voice their opinions \cite{chau2007mining}. Additionally, there are many other groups that are commonly targeted in cybercrimes, e.g., females, the disabled, and lesbian, gay, bisexual, and transgender (LGBT) individuals. Online harassment and threats are two of the common types of aggression against these groups \cite{lenhart2016dsri}.

Having introduced the ways in which to consider the group element to cybercrime, the body of this chapter seeks to critically examine it in further detail. First, it considers the platforms that are used by online groups, including Internet forums and mobile apps. Next, it examines the types of groupings present, including their actions and the factors that motivate crimes and draws heavily on case study examples arising from literature and the news. Then it builds on this foundation to analyze how criminal groups form and operate. This discussion encompasses issues of trust, motives, and means. The aim is to make these discussions pragmatic and provide useful insight into the group component of cybercrime, and issues such as interaction within criminal communities.

\section{Cybercrime and Online Groups}	

\subsection{Platforms Used by Online Groups: A Brief Look}	
Online groups and communities---or simply, people who interact via virtual environments---have existed since the dawn of the Internet. The first widespread groupings could be found on platforms such as Internet Relay Chat (IRC) and in chat-rooms on the once thriving AOL service. Since then, groups have spread to various other online services including social media services such as MySpace, Bebo, and Hi5, over the years. More recent social networks, including Facebook, host a number of groups on a range of diverse topics. The group aspect of Facebook is actually one of its most popular offerings with in excess of 1 billion users, and in December 2016, more than 10 billion comments and 25 billion likes \cite{friers2017fb}. Other networks, such as Twitter and Instagram allow groups to chat, but these are currently via direct messages as opposed to being more openly accessible (to join, use, etc.), as with Facebook. We do note here however, that Facebook does have several closed groups where access is strictly moderated, and may be based on demographics, interest, status, or employer.

Research has explored groups’ use of social media and, as alluded to above, they range from the benign to the more disruptive. For instance, groups have been used for teaching and improving writing \cite{yunus2012effectiveness}, but also for activist networks, be they associated with contemporary activism or collective action \cite{gerbaudo2012tweets,vromen2015young}. One significant challenge for social network platforms in this context has been maintaining the balance between freedom of speech (or excessive censorship) and the public good, particularly when considering those activist groups that may be viewed as extreme. This is a problem that social networks have struggled with for many years, and one that does not appear to be solvable through any simple or individual means.

Forums are another popular venue where groups form and interact. This platform functions analogous to a message board with posts sequentially added by date and/or time to a webpage. Some of the most well-known forums online are Reddit (the self-deemed ``Front page of the Internet''), CraigsList (a classified advertisements and discussion website), and 4Chan (an image-board website that allows anonymous posting). Positive uses of forums can be seen in activities such as support groups and those used for health advice \cite{cole2016health}, though negative uses are also abundant. Articles in research have even emphasized that Internet forums often act as an efficient and widely used tool for radical, extremist, and other ideologically ``sensitive'' groups and organizations to connect and inform on their agendas \cite{holtz2012analyzing}.

While cybercriminal and terrorist forums can be found on the open web, the most significant and devious are rife on the Dark Web. The Dark Web is the part of the web which exists on an encrypted network and can only be accessed using specific software and networks, such as Tor (or, The Onion Router) and I2P (the Invisible Internet Project). These services provide some level of anonymity hence their attraction to criminals. Dark Web forums and communications have been the focus of researchers for several years as they attempt to better understand how cybercriminals behave and act \cite{trendmicro2016dark}.

To complement online social media and forums, there are an increasing range of applications which allow groups to form and communicate. WhatsApp is one of the most popular of these applications, with around 1 billion users. This platform allows individual and group chat, and boasts secure messaging, a feature which has privacy advantages but is also heavily contested by governments and intelligence communities. Secure messaging in this context refers to WhatsApp’s use of full end-to-end encryption, which means that the only persons who can read messages (including photos, videos, files, etc.) are the sender and the intended recipients. As pointed out by WhatsApp, even they cannot see inside the messages \cite{whatsapp2017}.

A key reason why there is such heated deliberation around the services of WhatsApp and Telegram (a service similar to WhatsApp which also has end-to-end encrypted messaging), is because they may be seen as a safe space for terrorists and other criminals \cite{magdy2016safe}. There is no shortage of news articles and blogs which suggest this, as can be seen from the following story titles: ``Paris terrorists used WhatsApp and Telegram to plot attacks according to investigators'' \cite{billingtonj2015paris}, ``Inside the app that’s become ISIS' biggest propaganda machine'' \cite{englep2015app}, ``How Telegram Became The App Of Choice For ISIS'' \cite{robins2017telegram}, ``How terrorists use encrypted messaging apps to plot, recruit and attack'' \cite{hamill2017terror}, ``WhatsApp accused of giving terrorists `a secret place to hide' as it refuses to hand over London attacker’s messages'' \cite{rayner2017}  and ``Indian Govt May Ban WhatsApp Use In Country, As It Is Terrorist's Favourite App For Messaging'' \cite{dmello2018}.

Other apps and instant messaging services that authors (e.g., Magdy, 2016) have found that may be used by activist groups include WeChat (a China-based platform that has over 1 billion monthly active users), SureSpot (an open-source secure mobile messaging app that uses end-to-end encryption) and Kik (a messaging platform originating in Canada that has approximately 300 million users). To add to these, a topical study by TrendMicro of over 2,000 accounts that openly support terrorist groups has also found Wickr (an app that offers secure, ephemeral messaging) and Signal (an open source encrypted communications app) as preferred apps for these groups of individuals \cite{trendmicro2016dark}. As noted by Magdy \cite{magdy2016safe}, this range of apps may be used for different purposes but generally their popularity is driven by the fact that they allow faster, more personalized and secure communication.

\subsection{Groups as Perpetrators and Victims of Cybercrime}	
The platforms presented above have supported a range of activities pertaining to the group element of cybercrime. As discussed earlier, at least two approaches that could be explored are groups as the perpetrators and initiators of crime, or groups as the victims thereof, for instance, targeted demographics or minorities. Our aim in this section is to reflect on the group element of cybercrime more critically and identify the set of core group types. These would be beneficial to research and practice in the fields of study in cybercrime, cyberpsychology and criminology.

We begin this analysis with a consideration of the perpetrator’s perspective and thus, first look to understand how the public and literature perceive criminal groups. The three descriptions of criminal groupings that form the basis for our discussion are taken from the US Federal Bureau of Investigation, and the research works of Finckenauer and Voronin \cite{finckenauer2001threat} and Godson \cite{godson2003transnational}. We focus on organized criminal groups here as these are the most commonly discussed in the literature.

The US Federal Bureau of Investigation (FBI) considers the topic of transnational organized crime and the definition they ascribe to is:

\begin{quote}
``Those self-perpetuating associations of individuals who operate transnationally for the purpose of obtaining power, influence, and monetary and/or commercial gains, wholly or in part by illegal means, while protecting their activities through a pattern of corruption and/or violence, or while protecting their illegal activities through a transnational organizational structure and the exploitation of transnational commerce or communication mechanisms.'' \cite{FBI2016organcrime}
\end{quote}

\noindent From an academic perspective, Finckenauer and Voronin give insight into the group nature of crime through their definition of organized crime.

\begin{quote}
	``Organized crime is crime committed by criminal organizations whose existence has continuity over time and across crimes, and that use systematic violence and corruption to facilitate their criminal activities. These criminal organizations have varying capacities to inflict economic, physical, psychological, and societal harm. The greater their capacity to harm, the greater the danger they pose to society.'' \cite{finckenauer2001threat}
\end{quote}

\noindent Finally, Godson provides another academic definition on organized crime as he notes:

\begin{quote}
	``Organized crime refers to individuals and groups with ongoing working relationships who make their living primarily through activities that one or more states deem illegal and criminal. Organized crime can take a variety of institutional or organizational forms. This includes tight vertical hierarchies with lifelong commitments, as well as looser, more ephemeral, nonhierarchical relationships.'' \cite{godson2003transnational}
\end{quote}

Reflecting on these three definitions, we can begin to see some of the key features of criminal groups. For instance, there is the notion of continuity and group identity in the group (and member relationships) and criminal activities over time. This is particularly evident in the descriptions from the FBI and Finckenauer and Voronin. Motivation is another feature that stands out in the definitions, with influence and financial and commercial gain, acting as common reasons for group formation and crimes. Finckenauer and Voronin extend this point to highlight the generic aims of crimes; that is, inflicting economic, physical, psychological, and societal harms, but also the varying capabilities that criminal groups may possess in achieving such goals.

Godson touches on another important feature in terms of the various organizational forms that groups may take; for instance, they may be tightly bound or ephemeral and non-hierarchical. Group shape will likely depend on their nature and purpose, and the extent to which their activities will interest law enforcement. The FBI description is useful particularly because it emphasizes the transnational nature of criminal groups and their use of global communication channels, many similar to the platforms discussed earlier in this chapter and other chapters \cite{nurse2018cy}.

The reflection on criminal groups is crucial to the discussion on cybercrime for numerous reasons. In particular, there is almost certain to be many similarities between these groups considering that the Internet may be regarded as just another platform through which crime can occur. The various descriptions above can all be related in some way to cybercriminal groups. The main difference with cybercriminal groups is their focus on technology as a central means for interaction and criminal acts. Unlike traditional crime therefore, physical presence and power (including physical violence) is not as crucial, and technical means and skill tend to be more important. Furthermore, because of technology, cybercriminal groups can become transnational much more easily as they can meet and interact via the various platforms mentioned. Such interactions may be persistent or temporal depending on the nature of the crime. As we will discuss later, there is also the reality that with the Internet, forming groups of like-minded individuals is significantly easier than it is offline. There is less risk to group formation and persistence as well, given the ability online to mask one’s identity---these factors often combine to the advantage of criminals. Technology also means a wider availability of hacking platforms and tools, a reality that is predicted to increase in the future via a proliferation offensive tools \cite{williams2016future}.

Cybercrime groups have been of interest to researchers for some time, and therefore it is not surprising that articles have proposed ways to typify such groups. Possibly one of the most notable pieces of research on the topic is by Choo and Smith \cite{choo2008criminal}. They explore the exploitation of online systems by criminal groups and have defined three categories of such groups. The first category is that of traditional organized criminals who use technology to enhance terrestrial criminal activities. This includes crime syndicates and organized groups that specialize in everything from fraud and forgery to piracy and extortion from online gambling. Their aim is often to apply technology to expand and streamline operations.

Europol has carried out extensive work in the cybercrime space and have highlighted the prevalence of crime-as-a-service business models as a facilitator for traditional groups engaging in cybercrimes \cite{aikenmp2014}. Crime-as-a-service models, which can typically be found on underground Dark Web markets, allow criminals to purchase criminal services including acquiring botnets (or spam networks), launching denial-of-service attacks against specified targets, and customized malware development. As such, criminals can easily and quickly launch sophisticated cyber-attacks on groups or individuals of their choosing.

Organized cybercriminal groups are the second category identified by Choo and Smith and are said to be groups comprised of like-minded individuals working collectively towards a common goal. The Internet is a central enabler to such groups as it is the platform that they meet and plan activities; furthermore, their members may only be known to each other online. These are a few of the factors that distinguish these groups from traditional organized criminals which use technology to enable crimes. One example of such a group is the hacking group Lulzsec, where there are reports that their members never met in person, and were unaware of each other’s identities \cite{arthur2013lul}. Another recent example is the Carbanak cybercrime group, named after a piece of malware it used to access banking systems. The head of this group was the mastermind, and also technically talented enough to be able to identify software vulnerabilities and write malware to exploit them \cite{burgess2018}. According to reports, the head of the group also worked with three other gang members, who did not know each other and instead chatted online \cite{burgess2018}.

The last group category is that of ideologically and politically motivated cyber groups. This spans terrorist organizations and the full range of hacktivist groups. Choo and Smith make an intriguing point in their characterization of this category of groups. That is, that crimes often associated with organized criminal groups (e.g., scam and fraud schemes) are also crimes which terrorist groups engage in to raise funds for their ideological pursuits. A 2015 U.K. report showed that scamming and ransoms are high on the list of activities undertaken for terrorist financing \cite{HMTres2015}. Terrorist groups, e.g., ISIS, are widely known to engage in online activities, but particularly for plotting, recruiting, and claiming responsibility for attacks \cite{englep2015app,nouh2016towards,hamill2017terror}. Social media continues to be a favored platform for such groups, e.g., the role of Twitter in ``Tweeting the Jihad'' \cite{klausen2015tweeting}.

Hacktivists, or politically-motivated hackers, are also an increasingly popular grouping in this cybercriminal category. Such groups are known to carry out activities against governments and large corporations. Anonymous is one of the most well-known of these groups, given its attacks on the FBI and other sites \cite{peckham2012sites}. A key factor that makes Anonymous stand out potentially even more however, is its public-facing nature. There have been a variety of books published on Anonymous including Parmy Olson's We are Anonymous \cite{olson2013we} and Gabriella Coleman's Hacker, Hoaxer, Whistleblower, Spy: The Many Faces of Anonymous \cite{coleman2014hacker}.  Moreover, documentaries have been released on the workings and beliefs of its members---see We Are Legion: The Story of the Hacktivists.

Anonymous also maintains a significant presence on social media. At the time of writing, for instance, they appear to possess several Twitter profiles including @AnonyOps, @YourAnonNews, @YourAnonGlobal, @GroupAnon, @AnonPress, and @AnonyPress; the most popular being @YourAnonNews with more than 1.6 million followers. These various accounts hint to a core value of Anonymous, namely, the lack of central or hierarchical structure (as will be discussed in the third section of this chapter). This is clearly exemplified in the @GroupAnon tweet: “No, this is not the official \#Anonymous account. There is no official account. We have no central leadership. (Other than the FBI/NSA, joke)” made on 10:39 a.m. 18 Nov 2015.

While the three main groupings highlighted here are undoubtedly the core criminal networks, the authors of this chapter believe that there is another group, whose criminality is much more subjective, emerging in society today---individuals (often not criminals) who use technological means to motivate and organize acts that may be deemed dangerous or illegal. In Section 1, we presented one of these cases where there was a call to protest against President Trump’s January 2017 inauguration using a distributed denial-of-service attack (DDoS) on WhiteHouse.gov \cite{metzger2017ddos}. DDoS are regarded as criminal acts by many given that they are commonly used by hacker collectives to force legitimate websites offline.

The case of President Trump is interesting for many reasons. For example, there have been many rallies and protests against President Trump since he began his election campaign, several of which were organized online \cite{cbsnews2017behind}. While participating in rallies and protests is every citizen's right, the challenge of crime arises when these protests turn violent as they did in Portland, Oregon after the election and in Washington DC at the time of President Trump’s inauguration. In DC in particular, demonstrators set cars on fire and smashed shop and car windows \cite{longbottom2017protests}.

Civil action, organized via online networks and platforms, has also been witnessed in many other parts of the world prior to these U.S. instances. In the UK in 2011, thousands rioted following the death of local man Mark Duggan who was shot by the police; these riots led to mass looting and millions of pounds worth of property damage. It is said that the Blackberry Messenger app played a crucial role in the organizing the riots \cite{fuchs2012social} in enabling contagion and a group-mob mentality; Reicher \cite{reicher2001psychology} and Stott, Drury and Reicher \cite{stott2016role} provide further insight into the psychology of crowd dynamics broadly, and in the London riots, respectively. Facebook has also been used by activists as a platform for action and engagement with increased online activity found to often correlate with offline group actions \cite{nouh2015identifying}. In 2010’s Arab Spring, Facebook was used to spread the word of the revolution, and many believe that social media contributed to the liberation of those societies \cite{fuchs2012social}. These are all instances where technology and online interactions have contributed to offline unrest (be it positively or negatively motivated) and, in some instances, crime. There are many group processes at play in these instances, as there are offline. Establishing group identity and common goals plays a crucial role in bringing together individuals to create these groups.

In addition to the work by Choo and Smith \cite{choo2008criminal}, other articles that have sought to identify the types of cybercriminal group include McGuire \cite{mcguire2012organised} and Leukfeldt, Kleemans, and Stol \cite{leukfeldt2016typology}. McGuire suggests a typology of cybercrime groups with three main types. These are, groups that operate primarily online, those that combine online and offline activities, and groups that are predominantly offline but use online technologies as an enabler for crimes. This typology therefore closely matches up with the categorization of Choo and Smith. The research by Leukfeldt and colleagues adds another dimension to the analysis of cybercrime groups by considering them according to their characteristics. Specifically, they propose technology use (low-tech to high-tech) and the level of offender-victim interaction (no interaction to high interaction), while also noting the extent to which groups have local or international components. The benefit of such an approach is that it allows the correlation of characteristics, and in their case, the discovery of which types of network operate at which levels.

Having reflected on the perpetrator perspective of cybercrime and groups, we now consider the viewpoint of groups as the victims of online crime. While practically any demographic or characteristic can be used to target groups of individuals, some of the most common are those of race, religion, age, gender, and sexual-orientation. It is worth noting that these characteristics are not specially targeted in the online space but happen to be more openly targeted because of the illusion of attacker anonymity online. There are plenty of examples of groups that have formed online to preach hate towards persons of the characteristics highlighted. Chau and Xu \cite{chau2007mining} study one such type of hate group of ``anti-Blacks'' covering 820 bloggers on blog-hosting website, Xanga. A key finding from that research is that hate groups in the blogosphere may not tend to form into centralized organizations. The authors, however, do not eliminate the possibility that such online groups may prepare members for other extremist organizations such as the Ku Klux Klan, for instance.

Beyond race, religion is a significant factor in online victimization. A salient example of this victimization occurred after the Woolwich attack in May 2013 in the UK, where two Islamist terrorists brutally murdered a British soldier. In the days that followed, there were hundreds of messages on social networks containing hate speech directed against the Muslim community \cite{awan2014islamophobia}. Awan found that Muslims were demonized and vilified through negative comments, discrimination, physical threats, and online harassment. Other works have demonstrated this hate towards groups in online message boards as yet another example of how online platforms can be used to target people of certain faiths \cite{cleland2017islamophobia}.

While other groups (females, the disabled, and lesbian, gay, bisexual, and transgender (LGBT) individuals) are also the victim of online harassment \cite{lenhart2016dsri,chahal2016supporting}, youth are a particularly well considered area (by both academia and law enforcement) given their vulnerable nature. Kowalski and colleagues \cite{kowalski2014bullying} focus on the crime of cyberbullying among young people to provide a critical review of the existing body of cyberbullying research. Mitchell, Wolak, and Finkelhor \cite{mitchell2008blogs} also offer relevant insight that young Internet bloggers also were at an increased risk for online harassment. Furthermore, young individuals who interacted with people that they met online were at a higher likelihood of receiving online sexual solicitations. Population-based studies from other countries, e.g., Oksanen and Keipi \cite{oksanen2013young}, have supported these points and found that young people are generally more likely to be victims of cybercrime. A key novelty of their work is that they consider the risks of victimization that young people face online, to the problems they may face in the offline world.

Drawing on this analysis of the group component of cybercrime, Table \ref{label1} presents two main group types of groups: perpetrators and victims of criminal. The core subtypes of the former group are largely motivated by the work of Choo and Smith \cite{choo2008criminal}. To this has been added a new group focusing on citizens who use online technological means to mobilize and act. It is important to note that, in most instances, such action is not criminal and only in a few cases results in criminal acts (e.g., offline riots or looting). Furthermore, there may be arguments that this group is already accounted for in the ``Organized ideologically and politically motivated cyber groups.'' It is presented separately here due to its increasing importance in society (with the Arab Spring and the Trump protests arguably only the beginning of what is to come) and the difficulty in categorizing it, given it often borders on criminality.

\begin{table}[]
	\caption{The group element of cybercrime and its main types.}
		\label{label1}
	\begin{tabular}{|p{0.25\textwidth}|p{0.7\textwidth}|}
		\hline
		Main group types                                                 & Group subtypes \\
		\hline
	   Groups as perpetrators &    
\begin{enumerate}
	\item Traditional organized crime groups that use technology to enable crime.
	\item Traditional organized cybercriminal groups.
	\item Traditional organized ideologically and politically motivated cyber groups.
	\item Traditional citizen groups that use technology to mobilize and act.
\end{enumerate}
	        \\		\hline
		Groups as victims                            
		
                  &        
                  
\begin{enumerate}
	\item Race
	\item Age
	\item Disability
	\item Religion/belief
	\item Sex
	\item Sexual orientation
\end{enumerate}	\\	\hline
  
	\end{tabular}
\end{table}

With regards to the category of groups as victims, the subtypes listed have been studied in various articles before. The list included in Table \ref{label1} is based heavily on such works and instances of discrimination, victimization, and harassment found online. It is worth noting that these groups align broadly with the protected characteristics of the UK's Equality Act 2010 and similar legislation across the word. This emphasizes their significance in society more widely other than just in cyberspace. Over the next few years, there is expected to be a sharp growth in research into ``groups as victims'' online, particularly because of the difficulty that platforms such as Facebook and Twitter have in detecting and responding to online abuse and harassment.

\section{How Online Criminal Groups Form and Operate}	

With the main types of groups identified, this section narrows the focus to ``groups as perpetrators.'' It concentrates specifically on how cybercriminal groups form, engage, and operate.

\subsection{Group Formation and the Platforms and Networks That Enable It}	

Case studies suggest that within cybercriminal networks the importance of traditional central actors with the role of ``bridge builder''diminishes \cite{holt2014examining,motoyama2011analysis}. However, recent studies also show that such networks still have important social dependency relationships \cite{leukfeldt2016typology,leukfeldt2017cybercriminal,leukfeldt2017origin}. Research demonstrates that most of the networks have a (more or less) stable group of core members who commit crimes together over an extended period of time. The core members of these networks may know each other from the offline world and recruit only a few specialists through online meeting places. Other studies suggest that cybercriminal networks use offline social ties and, on occasion, online meeting places to come into contact with suitable co-offenders \cite{leukfeldt2017origin,leukfeldt2016typology,odinot2017organised}. Thus, the reality is that a minority of networks could be labeled as ad hoc networks that were forged in online meeting places to execute one-off cyber-attacks.

Social ties may be strongly clustered and limited to, for example, a region or country. Members of some cybercriminal networks are located in the same offline social cluster---even when executing cybercriminal attacks all over the world \cite{leukfeldt2017origin,leukfeldt2016typology,odinot2017organised}. Working with trusted acquaintances from the offline world could potentially have many advantages over working with potentially unreliable actors from all over the world who are only known by their online handle (pseudonym).

As with most situations, there are some exceptions to the common case where offenders are distributed across the Internet and not necessarily geographically located in one single place. The hacking group LulzSec is an example of this which is held to have been formed in private online chatrooms of the hacking collective Anonymous. Most notably, LulzSec members never met in the real world \cite{arthur2013lul}. From this example, it can be inferred that cybercrimes and cybercriminals, by their very informational, networked, and global nature, may go against the traditional model of socially and geographically rooted organized crime models. This pertains to the need to gather specialist skills; in particular, such groups tend to have a very detailed division of labor with specific skill sets across individuals. For instance, one person would provide the documents, another would buy credit card details, still another would create identities, and a fourth would provide the drop address \cite{rogers2006two}.

Furthermore, not all cybercriminals commit only cybercrimes. Studies suggest that cybercriminals are often also involved in all sorts of offline crimes \cite{leukfeldt2016typology,vanderbroek2016}. Yet, in the online world, distance, location, and time are no longer limiting factors. Compared to the offline world, it is relatively easy for offenders to be part of different criminal networks. For example, newcomers on forums are able to come into contact with existing members quickly and are able to reach a more central position relatively quickly.
To consider enabling platforms for criminal activities, the Internet has several criminal meeting places. Two examples are the forums and chat rooms where criminals meet to exchange information or make plans to carry out attacks. To a certain extent, forums can be regarded as platforms that facilitate the origin and growth of cybercriminal networks. Members of cybercriminal networks spend much of their time in criminal and non-criminal chat rooms and forums, where they meet like-minded people and build relationships. As mentioned, existing offline cultures, communities and social relationships also appear to be important in online forums \cite{ablon2014markets}.

Additionally, Leukfeldt, Kleemans et al. (2016) found that both social ties and online forums were used by cybercriminal networks to recruit new members. Four types of growth were identified in their work: 1) growth entirely through social contacts; 2) social contacts as a base and forums to recruit specialists; 3) forums as a base and social contacts to recruit local criminals; and 4) growth entirely through forums. Criminals would usually recruit through social ties and less through social contacts and use forums in order to find specialized enablers. An example of such a group is LulzSec. LulzSec’s members never met in the real world and were unaware of each other’s identities. Some were based in the US, and some in the UK, demonstrating the globalized nature of such groups.

Cybercriminals show a noticeable preference for carding forums. These are websites dedicated to the sharing of stolen credit card information as well as providing discussion boards in which members of the forum may share techniques used in obtaining credit card information. Using interaction data from three prominent carding forums---Shadowcrew, Cardersmarket, and Darkmarket---and drawing on theories from criminology, social psychology, economics, and network science, Yip, Webber, and Shadbolt \cite{yip2013forums} identified fundamental socio-economic mechanisms offered by carding forums: formal control and co-ordination, social networking, identity uncertainty mitigation, and quality uncertainty mitigation. Together, these mechanisms give rise to a sophisticated underground market regulatory system that facilitates underground trading over the Internet and thus drives the expansion of the underground crime economy. This demonstrates the robustness of carding forums and alludes to why they are favored by cybercriminals. Moreover, Holt and Lampke \cite{holt2010exploring} manually analyzed six forums and found that the dynamics of the stolen data markets are governed by key factors, including communications, price, quality, and service. This is intriguingly similar to legitimate markets.

To understand the cyber-threat landscape, it is also important to acknowledge the different ways that cybercriminal groups are organized. First, the cybercrime-as-a-service business model that drives criminal forums on the Dark Web provides the access to tools and services to people with little knowledge of cyber matters. Furthermore, the environment promotes exchange of information as well as ``learning kits.'' This trend is indicative of a growing cyber capability among these criminal groups as their knowledge expands and they exchange expertise. As some terrorist groups are reaching out to recruit in the Western world, they might be able to contact and attract appropriately skilled people for their hacking exploits \cite{aikenmp2014,ncsc2017cyberc}.

For the most organized and technically advanced groups, however, many of the services are carried out ``in-house'' as part of their own business model. For smaller groups or individual criminals, these services can be hired in one of many online criminal marketplaces. Most of these services like crime-as-a-service are openly advertised in criminal forums. As Richardson \cite{richardson2007csi} states, hackers have organized and shifted toward a ``professionalization'' of computer crimes. A few examples of such criminal forums are Dark Market, Carders Market, Shadowcrew, Carder.su, Darkode, GhostMarket, and the Silk Road.

To analyze the relationships among hackers more generally, it is often common to find a decentralized network structure. Network centralization describes a quality of a group and it indicates the extent to which a network is organized around one or more central points, such as a node or a centroid \cite{wasserman1994social,nouh2015identifying}. Previous research has shown that the Shadowcrew hackers, for example, were part of a decentralized network, although not everyone in this group had the same type of role or position \cite{lu2010social}. The network structure of this infamous hacker group was established using social network analysis methods. Leaders were identified using actor centrality measures (degree, betweenness, closeness, and eigenvector) and were found to be even more involved in thirteen smaller sub-groups \cite{lu2010social}. Shadowcrew had the three characteristics of a team as defined by Best and Luckenbill \cite{best1994organ}: 1) elaborate division of labor; 2) mutual participation; and 3) association. From this observation, the inference is that the members of cybercriminal groups do not necessarily have to be organized around one central point in order to still maintain a hierarchical structure.

In addition, the organization of crime online may often follow a different logic to the organization of crime offline. This is a dis-organized model of organization \cite{wall2007cybercrime}. Existing work identifies a ``dis-organized'' or distributed model of organization, rather than a hierarchical command and control structure of cybercrime \cite{wall2015dis}. Network technologies and associated social media are creating new forms of networked social relationships that act as the source of new criminal opportunities \cite{wall2007cybercrime} and crimes such as stalking, bullying, fraud, and sextortion.

Anonymous is an example of a group which does not strictly organize itself and has both swarm and hub characteristics. The fact that Anonymous has no leader makes it difficult to even comprehend its organizational structure \cite{nortonq2012anon}. The structure of Anonymous has been loosely described as ``a series of relationships'' with no membership fee or initiation. Anyone who wants to be a part of Anonymous---an Anon---can simply claim allegiance. Many Anonymous members considered themselves crusaders for justice. Publicly, Anonymous persists in claiming to be non-hierarchical \cite{kushner2014masked}.

Apart from collaborating and recruiting their members, it is also interesting to note that organizations operating on the Dark Web seem to also be attacking each other, and trying to prevail over their criminal competitors \cite{catakoglu2017attacks}. These attacks could be defacements aimed at subverting the business of another organization in order to promote a competitor website; attempts to spy on communications initiated to, and from, another organization, theft of confidential data from a disguised File Transfer Protocol (FTP) server, or manual attacks against the custom application running the underground forum. These activities demonstrate the tensions between groups as they participate in these various platforms and networks.

\subsection{Trust as a Factor for Cybercriminal Group Formation}	
The concept of trust within the human factors domain has focused largely on the user gaining trust as a result of specific website content, attributes, ease of use, and related consumer-centric acceptance models \cite{corritore2003line,nurse2011information}. Trust is an enabler of online engagement but also certain levels of trust are required when assessing what is being offered or accessed.

Supporting the growth of the Dark Web, and presumably the trust gained by participants to engage, are anonymity networks like Tor. In fact, it is a mandatory feature of a number of Dark Web forums that participants use Tor and agree to transact only through the use of virtual currencies, such as Bitcoin \cite{bradbury2014unveiling} and, increasingly, Monero (a virtual currency with a strong focus on privacy). The users of cybercrime marketplaces must trust that such environments will maintain their anonymity and will also follow through with the service communicated, e.g., provision of information on stolen credit cards. Ironically, the uniqueness of the trust environment for Dark Web participants and hosts appears to distil to the singular issue of preserving anonymity \cite{lacey2015s}. Integrity as a basis for trust in the Dark Web can encapsulate the overall integrity of the marketplace in maintaining anonymity of its users and hosts, which also connects to Mayer, Davis and Schoorman \cite{mayer1995integrative}, who observed that anonymity is a binding mutual interest for participants. Trust is dynamic, because it can build, diminish, and be removed at any point.

According to Falcone, Singh, and Tan \cite{falcone2001trust}, various different kinds of trust should be modeled, designed, and implemented when speaking about trust in cyber-societies: 1) trust in the environment and in the infrastructure (the socio-technical system); 2) trust in personal agents and in mediating agents; 3) trust in potential partners; 4) trust in information sources; and 5) trust in warrantors and authorities. Parts of these different kinds of trust have complementary relations with each other. The final kind---trust in a system and/or process---can be the result of various trust attributions to the different components. When an agent has to decide about whether to trust another agent in the perspective of a co-operative relationship, each must weigh the opportunities given by the positive results of a successful trust (benefits of trust) against the risks that the trust might be exploited and betrayed: this problem is known as the trust dilemma. The trust dilemma is the direct consequence of uncertainty---here, the intrinsic social uncertainty \cite{falcone2001trust}, and is similar to the social exchange principle engage in offline relationships to garner trust between one or more people \cite{thibaut1959}.

For all criminals, a balance must be made between remaining anonymous in order to remain unseen by law enforcement, and retaining certain aspects of identity in order to attract potential criminal collaborators \cite{lusthaus2012trust}. Online identities are the foundation of a cybercriminal’s reputation, which provides incentive to maintain that identity or a variation of it. At the same time, there is a competing incentive to change online names regularly in order to create a distance from past crimes. Reputation in some ways may be regarded as the ``currency'' that cybercriminals trade in on the Dark Web.

Gambetta’s (2009) contributions to both criminology and signaling theory expand the understanding of the ways criminals identify themselves to each other and signal trustworthiness in an otherwise untrustworthy environment. Specifically, when there is information asymmetry, it is in a signaler’s best interests to signal their trustworthiness, regardless of whether they actually are. Untrustworthy actors attempt to mimic the signals used by their trustworthy counterparts, and it is in the receiver’s best interest to differentiate between the two. Legitimate actors use signals that may be too costly for untrustworthy actors to replicate, which provides a potential way for receivers to interpret signals produced. To minimize the risk of harm, forums provide informal mechanisms that encourage trust between participants and sanction less reputable actors \cite{holt2015examining}. Other options also include having required reputation or history to enter closed online forums or to earn the status of ``trusted seller'' \cite{yip2013forums}.

Even with a system such as a carding forum that is capable of providing multiple channels for trust to develop, there is still room for mistrust \cite{mccarthy2001crime,chiles1996integrating}. In cases of mistrust, members of groups can be doxxed, such as the true identities of the members of the LulzSec gang that were made public, which ultimately led to the FBI arresting LulzSec leader Hector ``Sabu'' Monsegur \cite{brightp2012}. The interested reader is referred to the previous chapter for further information on doxxing and other common cybercrimes \cite{nurse2018cy}.
The Darkode forum, which had between 250-300 members, is another interesting case that operated very carefully and was very exclusive. Darkode administrators made sure prospective members were heavily vetted \cite{FBI2015ncrime}. Similar to practices used by the Mafia, a potential candidate for forum membership had to be sponsored by an existing member and sent a formal invitation to join. In response, the candidate had to post an online introduction---a resume---highlighting their past criminal activity, particular cyber skills, and potential contributions to the forum. The forum’s active members decided whether to approve applications, which showcases the importance of trust in the formation of cybercriminal groups.

\subsection{Group Operations, Their Motives and Means}	
Different organizations such as Anonymous, LulzSec, and the Ghost Security Group each illustrate quite different sets of offender motivations, levels of professionalism, and organization, but they also possess some similarities in terms of their organizing principles \cite{wall2015dis}. There may even be noteworthy patterns and motives across the groups linked to the motive, operating capability and attacks of the cybercriminals \cite{thornton2018malicious}.

A core dynamic of different groups appears to be based upon a reputational economy that binds the group together. As Wall \cite{wall2015dis} describes, when looking at the similarities of different groups, it is possible to identify that the key players seek the assistance of a broader group of participants who exist outside the central grouping, but within the idea frame (the crime-motive). These can help in solving problems related to the criminal activity being designed, built, or carried out. There may even be a further layer of individuals linked to the group who are outside the idea frame and who will give advice on specific issues. Sometimes individuals fall out of the information loop, or they are pushed out, or they leave, which makes the structure ephemeral. In most cases, the structure of the group is flat and lacks a hierarchical command and control form.

In brief, cybercriminals display common characteristics in that they often are fairly ephemeral and amorphous in terms of organization, and flex according to the demands and opportunities. They also seem to be self-contained in structure \cite{mcguire2012organised,yip2013forums}. They may regularly be driven by an individual or by a very small group, but not always, because the organizing principle is often like-mindedness with a central common idea or ethic. In Anonymous, for instance, each cell or subgrouping follows an idea frame (motive). There are not necessarily any relationships or even communications between cells outside the nucleus, just an identification and affiliation with the core idea. The interesting fact here is that this distributed type of organization does possess some similarities with the organization of many offline organized groups. They also reflect the United Nations Office on Drugs and Crime (UNODC) \cite{unodc2002global} organized crime group typologies.

One of the most interesting aspects of these communities is that of their characteristics and how they function. For instance, these individuals are likeminded and therefore have some shared culture, at least in the context of their actions. This culture includes values as well as intergroup dynamics. These include own-group perceptions, attitudes, and behaviors, as well as those towards another group. A cybercriminal’s social identity may be defined by group membership, as well as the general features that define the group and differentiate it from others \cite{hogg2000we}.

The most sophisticated cybercrime organizations are characterized by substantial functional specialization and divisions of labor \cite{broadhurst2014organizations}. The organization of cybercrime may also occur at a wider level and involve networks of individuals who meet and interact within online discussion forums and chat rooms. Some discussion forums function as ``virtual'' black markets that advertise, for example, stolen credit card numbers \cite{holt2010exploring}. A comparison of individual offenders and criminal organizations reveals that both possess impressive skills \cite{broadhurst2014organizations}. Odinot and colleagues \cite{odinot2017organised}  suggest the characteristics of offenders that are important in the offline world, such as age, physical health, and social behavior, are less important within cybercriminal networks. There are new types of offender not previously found among traditional organized criminal groups: those with an IT background, young offenders, and ill/disabled offenders.

Criminal organizations might also possess a variety of aims, including defiance of authority, freedom of information, sexual gratification of members, and technological challenge. However, the profit motive is more apparent in the organizational cases than with individual offenders, as are the activities undertaken by organizations operating under state auspices, specifically those involving espionage and offensive cyber operations.

While profiling cybercriminals of any type, there are specific common characteristics requiring investigation, such as technical know-how, personal traits, social characteristics, and motivating factors \cite{nurse2014understanding}. These have been derived from over 100 cases and exclusive reviews of pertinent literature regarding crimes. Often, the prime motivator for the majority of cybercriminals is not only easy profit, but also curiosity \cite{malenkovichs2012cyber}. Furthermore, in evaluating the motivation of cybercriminals, it is safe to state that some criminal action will be motivated by ``need'' \cite{maslow1954motivation} or by work and/or environment characteristics \cite{hunt1969new}. For example, different groups such as Anonymous, LulzSec, and the Ghost Security Group each illustrate quite different sets of offender motivations, levels of professionalism, and organization, but they also possess some similarities in terms of their organizing principles \cite{wall2015dis}.

In terms of motive, Shinder \cite{shinder2010profiling} lists monetary gain, emotion, political or religious beliefs, sexual impulses, or even boredom or the desire for ``a little fun.'' While these factors are obviously linked to traditional or real-world crime, what is not yet clear is whether cybercrime has the same associations or etiology. Critical in this regard is the understanding of motive: transition from initial motive to sustaining motive, overlapping motives, and the prediction of evolving motives, along with an understanding of primary and secondary gains.

For example, a hacker becoming part of a community of like-minded persons involves a subcultural aspect inherent within creating online relationships that allow a hacker to express themselves \cite{bossler2012general}. This subculture might be characterized by the perception that committing cybercrimes is something normal. Within a group there will exist some resistance to perform immoral activities, while others with a lower moral threshold may opt or enlist to perform them to increase their benefit \cite{atkinsons2015}. It is of note that cybercriminals will protect the infrastructure rather than destroying it to keep making money from the persons and/or networks that they have compromised \cite{aikenmp2014}.

Perry \& Olsson \cite{perry2009cyberhate} found that the Web created a new common space that fostered a ``collective identity'' for previously fractured hate groups, strengthening their domestic presence in counties such as the US, Germany, and Sweden. McDevitt, Levin, and Bennett \cite{mcdevitt2002hate} identified broad categories of hate crime offenders: 1) thrill offenders---those who commit their crimes for the excitement or the thrill; 2) defensive offenders---those who view themselves as defending their ``turf''; 3) mission offenders---those whose life’s mission is to clear the world of groups they consider evil or inferior; and 4) retaliatory offenders—those who engage in retaliatory violence. Therefore, the motives of these groups define the way they operate.

Models of small group dynamics suggest how conformism, the influence of extremist ideologies on moving people to more extreme attitudes, disinhibition, and the yearning for group acceptance can all conspire to drive a person to commit acts of hate crime \cite{rieker1997ethnozentrismus}. Hate crimes can also be committed due to psychological, social-psychological, historical-cultural, sociological, economic, and political reasons \cite{green2001hate}.

LulzSec, Anonymous, and the Ghost Security Group offer useful practical examples. In the case of the first, the intention appeared to be gaining attention, embarrassing website owners, and ridiculing security measures  \cite{arthur2013lul}. For Anonymous-affiliated activists, perhaps the highest profile was their work under the banner ``Operation Isis,'' or \#OpISIS, which consisted largely of finding Twitter feeds that supported the ISIS terrorist group (and were often used to distribute propaganda and share news releases) and reporting them to Twitter so that they could be shut down \cite{griffin2015paris}. Lastly, Ghost Security Group also engaged in similar targeting of jihadists by monitoring suspected ISIS Twitter accounts and infiltrating militant message boards to find information, which they would then pass along to law enforcement \cite{bbc2015ghost}. These actions could be considered noble and of benefit to society, therefore hinting at the varying motives and values of such groups.

However, there are several instances to the contrary, e.g., under the banner \#OpTrump Anonymous targeted Donald Trump before he was elected president. The attack led to temporary shutdowns of Mr. Trump’s website and alleged hacks of his voicemail \cite{griffin2017anon}. It is worth reiterating that announcing that their next target would be the Trump campaign set off the most heated debate yet within the movement. Many disavowed the anti-Trump operation as being counter to Anonymous’ tradition of not taking sides in political contests \cite{woolf2016anti}. These conflicting aims are not surprising, given the dispersed nature of this cybercrime group. More importantly, it provides a perfect illustration of the context and reality of such online criminal groups and generally issues related to the group element of cybercrime.

\section{Conclusion}	
\label{sec:conclusion}
This chapter reflected on the group element of cybercrime to develop a better understanding of how groups may be perpetrators as well as targets of online crime. It provided an up-to-date analysis of the various online platforms used by cybercriminals as well as examined how these malevolent groups form, how their members develop trust in each other, and the motives that drive a group’s success and actions. In addition to elucidating these often-undefined aspects in research, it also presented a characterization of the group element of cybercrime and its main types, including newly emerging group types. The current research forms the basis for a more thorough understanding of online criminal groups, and thereby encourages further discussions on how they might be unraveled and potentially even thwarted.

\bibliographystyle{splncs03}
\bibliography{bibliosocs}

\end{document}